\documentclass[12pt,preprint]{aastex}

\shorttitle{Fe Fluorescence in the X-ray Spectrum of HR~9024}
\shortauthors{Testa et al.}

\def \rstar {$R_{\star}$}

\def \ne  {$n_{\rm e}$}

\def \ta  {Paper~I}

\def \hr       {HR~9024}
\def \cha      {{\em Chandra}}

\def \hetgs    {{\sc hetgs}}

\def \heg     {{\sc heg}}

\begin{document}
\title{Geometry Diagnostics of a Stellar Flare from Fluorescent X-rays}
\author{Paola Testa\altaffilmark{1}, Jeremy J.\ Drake \altaffilmark{2}, 
Barbara Ercolano\altaffilmark{2}, Fabio Reale\altaffilmark{3,4}, 
David P.\ Huenemoerder\altaffilmark{1}, Laura Affer\altaffilmark{3,4},
 Giuseppina Micela\altaffilmark{4}, David Garcia-Alvarez\altaffilmark{2,5}}
\altaffiltext{1}{Massachusetts Institute of Technology, Kavli 
Institute for Astrophysics and Space Research, 70 Vassar street, 
Cambridge, MA 02139, USA; testa@space.mit.edu}
\altaffiltext{2}{Smithsonian Astrophysical Observatory, MS
3, 60 Garden Street, Cambridge, MA 02138, USA}
\altaffiltext{3}{Dipartimento di Scienze Fisiche \& Astronomiche, 
Universit\`a di Palermo Piazza del Parlamento 
1, 90134 Palermo, Italy}
\altaffiltext{4}{INAF - Osservatorio Astronomico di Palermo,
Piazza del Parlamento 1, 90134 Palermo, Italy}
\altaffiltext{5}{Imperial College London, Blackett Laboratory,
Prince Consort Rd, London, SW7 2AZ, UK}

\begin{abstract}

We present evidence of Fe fluorescent emission in the \cha\ \hetgs\ 
spectrum of the single G-type giant \hr\ during a large flare. 
In analogy to solar X-ray observations, we interpret the observed
Fe~K$\alpha$ line as being produced by illumination of the photosphere 
by ionizing coronal X-rays, in which case, for a given Fe photospheric
abundance, its intensity depends on the height of the X-ray source. 
The \hetgs\ observations, together with 3D Monte Carlo calculations 
to model the fluorescence emission, are used to obtain a direct
geometric constraint on the scale height of the flaring coronal
plasma.  
We compute the Fe fluorescent emission induced by the emission of 
a single flaring coronal loop which well reproduces the observed X-ray
temporal and spectral properties according to a detailed hydrodynamic
modeling.  The predicted Fe fluorescent emission is in good agreement 
with the observed value within observational uncertainties, pointing 
to a scale height $\lesssim 0.3$\rstar.  
Comparison of the \hr\ flare with that recently observed on II~Peg by 
{\em Swift} indicates the latter is consistent with excitation by 
X-ray photoionization.

\end{abstract}
\keywords{hydrodynamics --- plasmas --- stars: coronae --- X-rays: stars }

\section{Introduction}
\label{s:intro}

Since the early 1970's, spatially resolved observations of the solar
corona have revealed a high degree of structuring over all scales,
from of order of the solar radius down to instrumental resolution limits
(e.g., \citealt{Vaiana73,Vaiana73b}).  This spatial structuring is
intimately linked to the characteristics of magnetic field generation
and interaction, and to plasma heating mechanisms.  On other stars,
especially very active ones (with X-ray luminosity $L_{\rm X}$ up to 
$10^4 \times L_{\rm X \odot}$), coronal structure and its relation 
with stellar parameters such as mass, rotation and evolutionary 
phase, remains very uncertain.

Techniques used to date to investigate the morphology of stellar coronae
comprise rotational modulation \citep[e.g.,][]{MarinoL03,Flaccomio05},
flare modeling (e.g., \citealt{Reale04}; \citealt{Testa07a}, hereafter
\ta), eclipse mapping \citep[e.g.,][]{White90}, spectroscopic density
and radiation field diagnostics \citep[e.g.,][]{Testa04b,Ness04},
resonance scattering \citep[e.g.,][]{Testa04a,Matranga05,Testa07b},
velocity modulation
\citep[e.g.,][]{Brickhouse01,Chung04,Hussain05,Huenemoerder06},
and simultaneous Doppler imaging and X-ray spectroscopy \citep{Hussain07}.

The fluorescent iron line at $\sim$6.4~keV \footnote{The Fe
  fluorescent feature actually consists of two components, at 6.391
  and 6.404~keV (for Fe\,{\sc i}), which are unresolved by present
  instruments, therefore hereafter we will refer to the feature as to
  a single emission line.}  presents a further, potentially powerful
diagnostic of stellar coronal geometry.  Extensively used in study of
active galactic nuclei and X-ray binaries (see e.g.\ reviews
\citealt{George91,Reynolds03}), and often seen on the Sun during
flares \citep{Culhane81,Parmar84,Tanaka84,Zarro92,Phillips94}, the
line is produced by electron cascade after one of the two K-shell
($n$=1) electrons of an iron atom (or ion) is ejected following
photoelectric absorption of X-rays.  In the solar and stellar case,
for a given irradiating flare spectrum the line strength depends
essentially on three parameters: the flare height, heliocentric angle,
and photospheric Fe abundance (e.g., \citealt{Bai79}; 
\citealt{Drake07b}, hereafter D07b).
If it can be observed in stars, Fe~K fluorescence therefore presents a
means of estimating flare and coronal scale height.

Fe fluorescence has now been detected on a number of pre-main sequence
stars with disks \citep[e.g.,][]{Tsujimoto05,Favata05}, where 
it is most likely produced by coronal X-ray irradiation of the cold 
disk material.  Here we present evidence for {\em photospheric} Fe
fluorescent emission in the \cha\ \hetgs\ spectrum of the X-ray active
single G1 giant \hr\ (see \ta\ and references therein for a
description of the characteristics of the target).  The observations
and analysis are briefly described in \S\ref{s:obs}. In
\S\ref{s:model} we present 3D Monte Carlo calculations of the Fe
K$\alpha$ fluorescence.  The results are presented in
\S\ref{s:results}, where we derive an estimate for the coronal scale
height and compare it with the prediction of the loop hydrodynamic
model that succeeds in describing the observed flaring spectrum (\ta).
To our knowledge, the \hr\ {\em Chandra} spectrum represents only the
second detection of Fe photospheric fluorescence.  Very recently, the
line was observed by the {\em Swift} X-ray Telescope during a
``superflare'' on the RS~CVn binary system II~Peg by \citet{Osten07},
who ascribed the excitation mechanism to electron impact ionization of
photospheric Fe instead of photoionization.  We discuss our result and
the II~Peg observations in \S\ref{s:discuss}, and show that the latter
is also consistent with, and more plausibly associated with,
photoionization than electron impact.

\section{Observations and Data Analysis}
\label{s:obs}

\cha\ High Energy Transmission Grating \citep{hetg05} 
observations of \hr\ (ObsID 1892) were discussed in detail by \ta\ 
that analyzed the X-ray spectra and variability of \hr\ and 
constructed a flaring loop hydrodynamic model that successfully 
reproduced the observations.  We analyse the same data here, using 
the PINTofALE\footnote{http://hea-www.harvard.edu/PINTofALE}
IDL\footnote{Interactive Data Language, Research Systems Inc.}
software \citep{PoA}, and the Interactive Spectral Interpretation
System (ISIS\footnote{ISIS is available at http://space.mit.edu/cxc/isis/}) 
version 1.4.2 \citep{Houck00}.

\begin{figure}[!hb]
\begin{center}
\epsscale{.6}
\plotone{f1a.ps}
\plotone{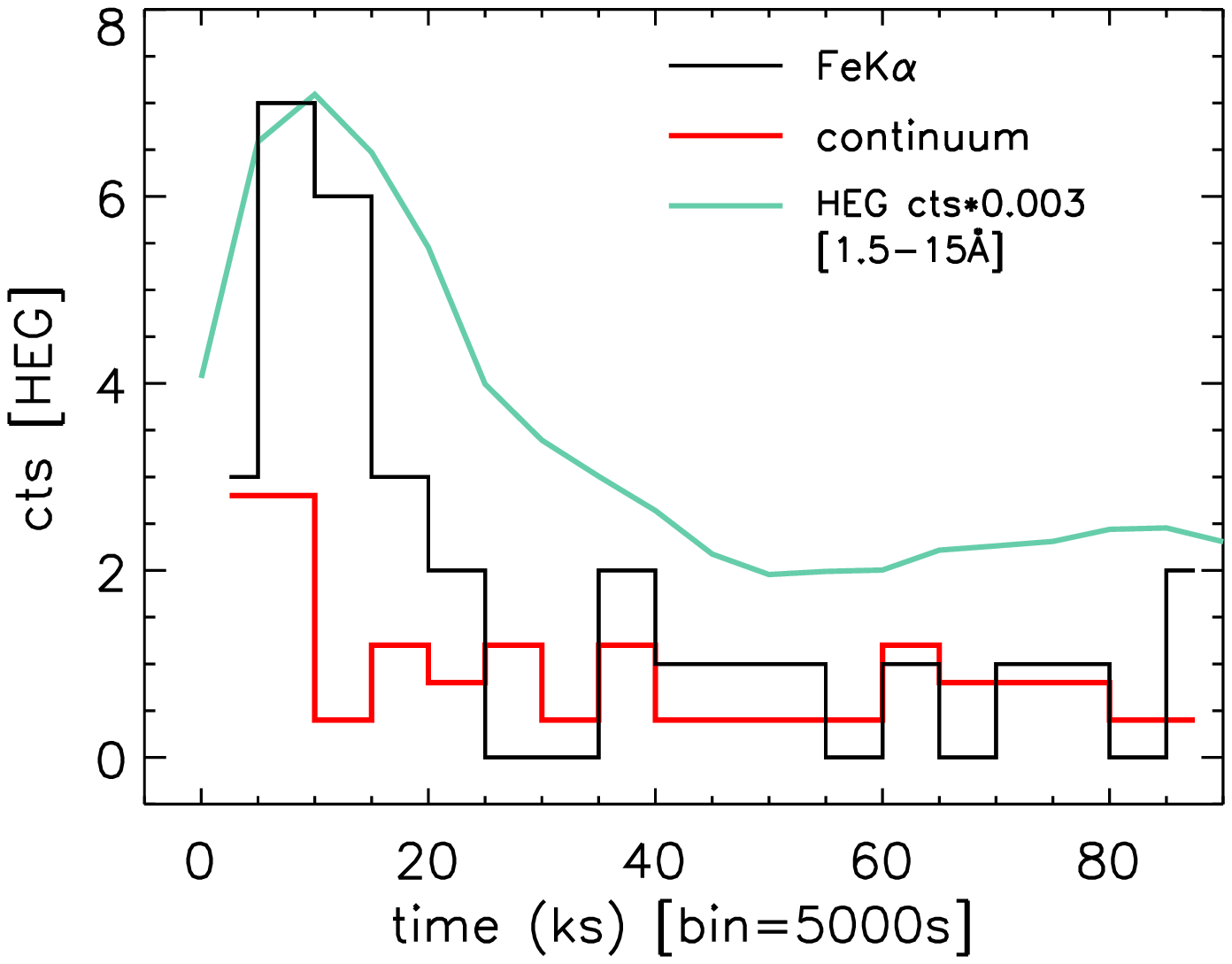}
\epsscale{1}
\caption{{\em Top:} \heg\ spectrum of HR~9024 during the flare
	(first 40~ks of the observation)
	in the wavelength range containing the Fe\,{\sc xxv} 
	coronal emission ($\sim 1.85$\AA) and the Fe K$\alpha$ 
	emission ($\sim 1.94$\AA), with the best fit model. 
	{\em Bottom:} Comparison of lightcurve of the K$\alpha$ 
	emission (black) with the adjacent continuum background 
	emission (red); the continuum is extracted on a wavelength 
	region (1.912-1.932\AA + 1.948-1.968\AA) that is 2.5 times 
	larger than the wavelength extraction region used for the 
	Fe K$\alpha$, and then appropriately scaled. 
	We also plot the scaled lightcurve of dispersed photons 
	integrated in the \heg\ band (green curve).
	A temporal binsize of 5~ks is used for all lightcurves.
	We plot counts (scaled by 0.003 in the case of the 
	integrated \heg\ band), therefore the associated 
	uncertainties correspond to the Poisson errors.
	\label{fig:spectra}}
\end{center}
\end{figure}

The \heg\ flare spectrum (first 40~ks of the observation) in the 
wavelength region with the Fe~K$\alpha$ emission is illustrated 
in Figure~\ref{fig:spectra} ({\em top panel}), together with a 
comparison between the evolution of the fluorescent and adjacent 
continuum emission ({\em bottom panel}).  
A Kolmogorov-Smirnov test applied to the lightcurves of Fe K$\alpha$ 
and the nearby continuum indicates that they are significantly 
different during the flare ($P < 10^{-10}$), but $P \sim 0.94$ during 
the rest of the observation.
The Fe K$\alpha$ line flux was measured with ISIS by convolving a
Gaussian profile with the instrumental response, using the 
continuum predicted by an isothermal spectral model described 
in \S\ref{s:model} (and derived in \ta).
We obtained an Fe K$\alpha$ line flux of $1.7 \times
10^{-5}$~ph~cm$^{-2}$~s$^{-1}$, with a corresponding 90\% confidence
range $0.8 - 2.8 \times 10^{-5}$~ph~cm$^{-2}$~s$^{-1}$.

\section{Fluorescence Modeling}
\label{s:model}

The basic physics of Fe fluorescent line production on the Sun has 
been investigated through detailed Monte Carlo calculations of the 
interaction between a hard X-ray continuum ($E > 7.11$~keV) and 
dense, neutral photospheric material by \citealt{Bai79},
considering a thermal continuum and taking into account the curvature 
of the solar photosphere.  The problem has been further discussed in
detail in the stellar context by D07b.

\begin{figure}[!ht]
\begin{center}
\plotone{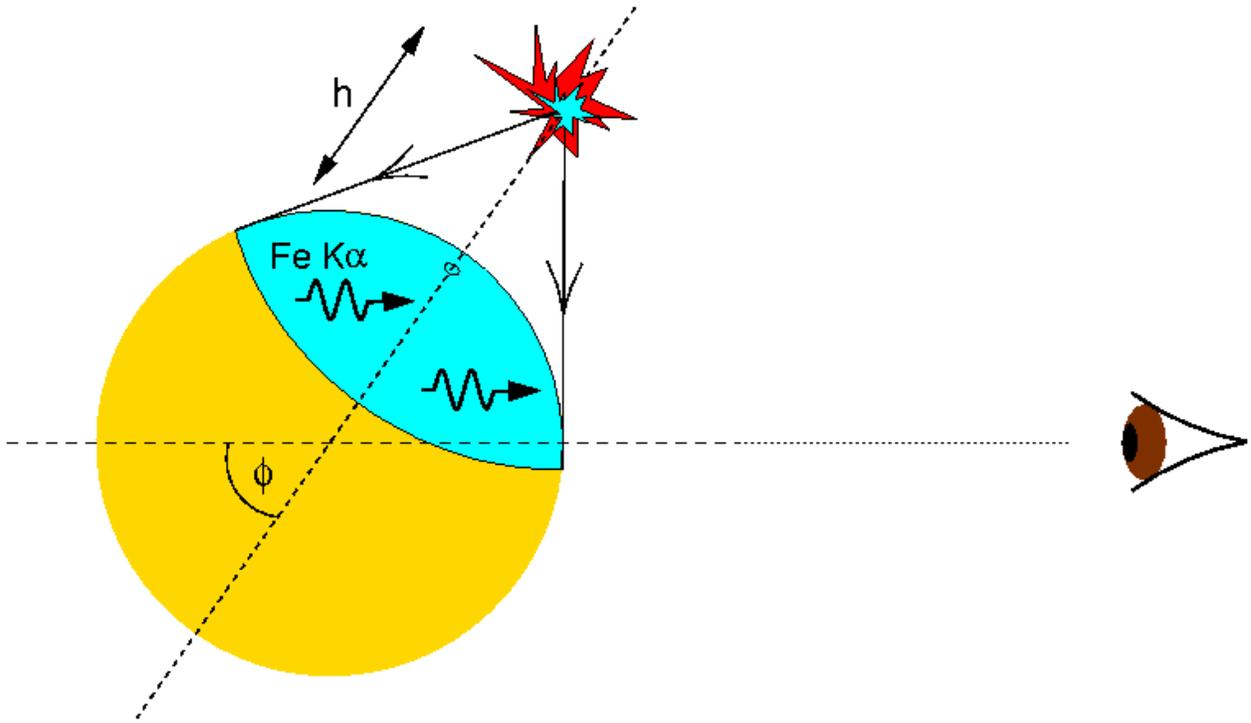}
\caption{Geometry for fluorescent excitation of photospheric
	Fe K$\alpha$ by coronal X-ray emitted by a source located
	at height $h$ above the photosphere. $\phi$ is the 
	angle of inclination with respect to the line-of-sight.
	\label{fig:geom}}
\end{center}
\end{figure}

We used a modified version of the 3D radiative transfer code MOCASSIN
\citep{Ercolano03,Ercolano07,Drake07}, to carry out Monte Carlo 
calculations to follow the Fe K$\alpha$ photons resulting from X-ray 
absorption and to determine the emerging Fe K$\alpha$ emission as a 
function of the flare geometry parameters illustrated in 
Figure~\ref{fig:geom}.  
Here, our method differs from the calculations of \cite{Bai79} in that 
we can use any given input fluorescing spectrum, whereas \cite{Bai79}
adopted an analytical expression for the thermal bremsstrahlung
continuum and computed a grid for T in the range 0.5-5~keV.  Detailed
descriptions of the methods used for the fluorescence calculations are
presented by D07b.

We note that the fluorescence efficiency, $\mathcal{E}$, defined as
the ratio of the Fe~K$\alpha$ flux to one-half the integral of the
X-ray flux above 7.11~keV ($\mathcal{E}=I({\rm
  FeK}\alpha)/[I_c(E>7.11~{\rm keV})/2]$) is only very weakly
dependent on the photospheric metallicity, [M/H] (D07b).  For \hr\ no
accurate photospheric abundances were available, and we therefore
obtained optical spectra with the high resolution ($R=86000$)
spectrograph SARG at the Telescopio Nazionale Galileo, and derived a
near-solar photospheric Fe abundance ${\rm [Fe/H]} = -0.2 \pm 0.3$
(L.\ Affer et al.\ in prep.).  
Based on the calculations of D07b, we
find an uncertainty in $\mathcal{E}$ of $\pm 25$\%\ 
for a metallicity uncertainty range [M/H]$=-0.5$--$+0.1$,
including allowance for an overabundance of $\alpha$-elements with
respect to Fe at the low metallicity end of 0.2-0.3~dex
\citep[e.g.][]{Bensby03}.

\begin{figure}[!ht]
\begin{center}
\plotone{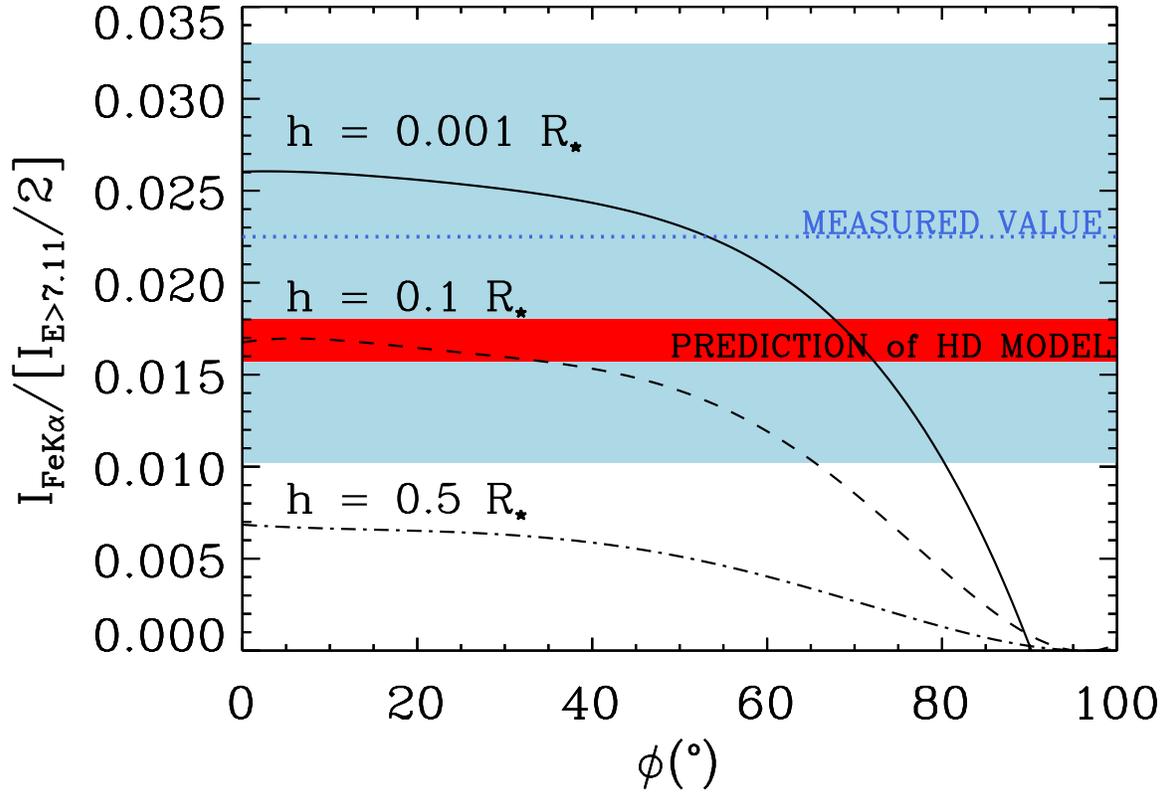}
\caption{Theoretical curves of fluorescence efficiency (black curves), 
	obtained using the MOCASSIN code, as a function of the angle 
	of inclination with respect to the line-of-sight, $\phi$, for 
	different heights of the hard X-ray source: $h= 0.001 R_\star, 
	0.1 R_\star, 0.5 R_\star$. We superimpose the measured value 
	(blue dotted line) with the corresponding 90\% confidence 
	range (shaded light blue area), and the prediction 
	of the hydrodynamic loop model (red shaded area).
	\label{fig:mod_obs}}
\end{center}
\end{figure}

In order to compute $\mathcal{E}$ we used a fluorescing spectral model
with $\log {\rm T [K]} = 7.8$,  and $\log {\rm EM [cm^{-3}]} = 54.9$,
based on the values found from the analysis of the continuum emission
in the early phases of the flare (see Fig.~6 of \ta).  This was
preferred over a spectrum computed using the emission measure
distribution, $EM(T)$, derived for the flare as a whole (rise + peak +
decay; see Figs.~3 and 6 of \ta) because most of the Fe K$\alpha$ 
emission is detected in the rise, peak, and early decay phases: 
the use of the whole flare $EM(T)$ would slightly underestimate the 
hard X-ray flux.  Values of $\mathcal{E}$ were calculated assuming a 
point source for the flare and for a range of flare heights, $h$.
The results obtained for a range of flare heights and line-of-sight 
viewing angles are shown in Figure~\ref{fig:mod_obs}.

As expected, the model fluorescent efficiency declines strongly for 
larger values of $h$, the height of the flaring source.  This occurs 
because of the $1/h^2$ dilution of the irradiating surface flux, and 
because photon incidence angles become larger, causing the average 
depth where the photoionization takes place to increase.  
For the line-of-sight viewing angle $\phi \lesssim 60^\circ$, the 
fluorescence efficiency depends primarily upon $h$ but not on $\phi$; 
for $\phi>60^\circ$, the efficiency rapidly decreases\footnote{The 
results are qualitatively similar also for the case of the semi-infinite 
slab in \cite{George91}.}.

In \ta\ we modeled the observed flare using a hydrodynamic
treatment of the evolution of the flaring plasma within the
confines of a single long loop.
The hydrodynamic model provides us with the physical conditions 
(in particular T and \ne) of each plasma volume element
along the loop, allowing us to compute the expected Fe K$\alpha$
emission for a direct comparison with the measured fluorescence
efficiency.  In order to calculate the $\mathcal{E}$ induced by
the emission of the flaring loop as resulting from
the hydrodynamic modeling we proceeded as follows: (1) we
synthesized hard X-ray spectra (7-50~keV) for isothermal plasmas for a
grid of temperatures covering a large T range ($\log T =5,8.5$); (2)
we divided the half-loop in four different segments and computed hard 
X-ray spectrum of each segment by adding the spectra of all
volume elements (each characterized by a temperature and emission
measure) within the segment, integrating over the 40~ks of the flare; 
(3) we computed the expected fluorescence
emission induced by each segment, depending on its input spectrum and
corresponding height, and computed the overall Fe K$\alpha$ emission,
and therefore the predicted $\mathcal{E}$.  Splitting the loop in
segments linearly spaced in $h$ (i.e., 
$h_i =\{0.125,0.375,0.625,0.875\} \times h$), assuming that the loop
is perpendicular to the stellar surface, we obtain 
$\mathcal{E} ({\rm HD~mod}) = 0.016$. 
We also computed the efficiency using a logarithmic spacing 
($h_i = \{0.0065,0.028,0.118,0.504\} \times h$), and we find 
$\mathcal{E} ({\rm HD~mod}) = 0.018$ only slightly different from the
former value.

\section{Results}
\label{s:results}

To compare the observed Fe~K$\alpha$ flux with model values, we used
the same spectral model based on the observed continuum described
above.  The one-half integrated X-ray flux above 7.11~keV  
(up to 50~keV) as computed from this model is
$7.5 \times 10^{-4}$~ph~cm$^{-2}$~s$^{-1}$, which, when combined with
the observed Fe~K$\alpha$ flux, yields a value for the observed
fluorescence efficiency $\mathcal{E} = 0.022~[0.010-0.033]$.
We note that at 50~keV the flux is already several orders 
of magnitude lower than at 7.11~keV, therefore the efficiency 
does not depend significantly on this upper bound value used for 
the energy range.
In Figure~\ref{fig:mod_obs} the measured value, with the 90\%\ 
confidence range, is represented as a shaded area superimposed to 
the theoretical curves derived in \S\ref{s:model}.  

The comparison between observed and model values of $\mathcal{E}$ 
implies for the scale height an upper limit of $\sim 0.3$\rstar,
where \rstar$=13.6 R_{\odot}$ \citep{Singh96a}.
This result compares well with the hydrodynamic 
modeling that finds a loop semi-length $L \sim 0.5$\rstar\ (\ta),
corresponding to a loop apex at $h = L \times 2/\pi =$
\rstar$/\pi \sim 0.32$\rstar\ (assuming that the loop is 
perpendicular to the stellar surface, otherwise $h < 0.32$\rstar). 
The efficiency expected on
the basis of the hydrodynamic model itself corresponds to an ``effective
scale height'' $h \sim 0.1$\rstar ---somewhat lower that the maximum
height of the loop model.  This can be understood on the basis
of the extreme characteristics of the flaring loop in \hr, which is 
much hotter than typical solar flares: for solar flares, typically
reaching temperatures $\gtrsim 10^7$~K, the hard X-ray photons
above the 7.11~keV threshold will be emitted almost uniquely by the 
hotter loop apex; instead, in the flaring loop on \hr, with 
peak temperature of the order of $10^8$~K, the lower coronal plasma
will still be effective in producing hard X-ray photons because of
its elevated temperature (of several $10^7$~K) and density.

\section{Discussion and conclusions}
\label{s:discuss}

We have presented the first detailed analysis of observed photospheric
fluorescence to probe the geometric properties of stellar coronal
emission.  For the flare on \hr\ we find the observed Fe~K$\alpha$ 
intensity completely consistent with production through photospheric 
irradiation by flare X-rays.  Our estimate for the flare scale height 
of $h \lesssim 0.3$\rstar\ provides a cross-check for the results of 
the hydrodynamic modeling of the flare observed on \hr\ that can be
satisfactorily reproduced by a single flaring loop of semi-length 
$L =0.5$\rstar.  The agreement between the observed $\mathcal{E}$ and
the value computed directly from this model provides further 
confidence both in the hydrodynamic modeling approach and X-ray 
fluorescence as the excitation mechanism for the Fe~K$\alpha$ line.
We note that our model based on \cha\ data might underestimate
the high energy continuum flux because we cannot detect any additional  
(either thermal or non-thermal) emission possibly present at 
energies outside the \cha\ band.
However, if additional high energy emission were present, the higher  
flux above 7.11~keV would make the efficiency lower bringing it  
in even better agreement with the predictions of the model. 

As noted in \S\ref{s:intro}, Fe K$\alpha$ emission has been 
extensively observed in solar X-ray spectra during flares
\citep{Culhane81,Parmar84,Tanaka84,Zarro92,Phillips94}.  Two main
production mechanisms were initially explored to explain the line:
electron impact from a non-thermal electron population, and X-ray
photoionization.  The observed Fe K$\alpha$ evolution with respect to
the thermal and non-thermal emission, and the observed center-to-limb
variations seen on the Sun, strongly support the X-ray fluorescence
mechanism \citep{Culhane81,Parmar84,Tanaka84}, although in rare cases
there is indication that electron impact might contribute during the
early impulsive phase \citep[e.g.,][]{Emslie86}.
\citet{Ballantyne03} have also shown that Fe~K production in
accretion disks by non-thermal electron bombardment is extremely
unlikely, and requires 2-4 orders of magnitude greater energy
dissipation in the electron beam than is required for an X-ray
photoionization source.  
In this context, the recent observations of an extremely large 
Fe~K$\alpha$ equivalent width in the PMS star V1486~Ori 
\citep{Czesla07}, and variability in Fe~K$\alpha$ emission 
uncorrelated with the observed X-ray continuum on Elias 29 
\citep{Giardino07}, seem to challenge the photoionization excitation 
mechanism.  However, D07b note that fluorescence from PMS disks 
excited by X-rays originating from the unseen stellar hemisphere 
will also be observed: in $\leq 50$\%\ of cases, Fe~K$\alpha$ can 
then appear anomalous when compared only to the {\em observed} thermal 
continuum.

To date, the {\em Swift} observation of II~Peg \citep{Osten07} is the
only other case where Fe~K$\alpha$ emission has been observed for a
star (other than the Sun) lacking substantial circumstellar material.
\cite{Osten07} dismissed X-ray fluorescence as the excitation
mechanism based on arguments for fluorescence in an optically-thin
medium, and instead attributed the Fe K$\alpha$ to electron impact
with non-thermal electrons.  As noted in earlier discussions
\citep[e.g.][]{Parmar84,Ballantyne03}, one major problem for
an electron impact excitation mechanism is the inefficiency of this
process.  Extremely large energies must be dissipated in accelerated
electron beams to produce continuum and fluorescence emission that
might be observed on a star.  Indeed, \citet{Osten07} showed that the
hard X-ray continuum observed in the II~Peg flare would require an
energy of $3 \times 10^{40}$~ergs in non-thermal electrons if
interpreted in terms of thick-target bremsstrahlung.  Over the time
of the duration of the flare, this corresponds to an energy
dissipation rate in accelerated electrons of more than 100 times the
stellar bolometric luminosity.  The hard X-rays are also interpreted
as lasting for timescales comparable with the flare soft X-ray
emission, at variance with the Sun where this non-thermal component,
when present, is generally impulsive. 
Regardless of the true nature of the hard X-ray flux, the equivalent 
width measured from the II~Peg {\em Swift} spectra ranges
between 18 and 60~eV, corresponding to efficiency values $\mathcal{E}$
between 1\% and 2\%.  These values are well within the range found
from the theoretical calculations presented here, and comparable with
the measured value for \hr.  Since the non-thermal hard X-ray
component in the II~Peg flare is not well-constrained by the 
{\em Swift} data and could also be explained by thermal emission, the
observed Fe~K$\alpha$ line seems more easily explained as arising
from photoionization by flare X-rays 
(see also the discussion of D07b).

\begin{acknowledgements}
PT and DPH were supported by SAO contract SV3-73016 to MIT for support 
of CXC, which is operated by SAO for and on behalf of NASA under 
contract NAS8-03060.  JJD was supported by the CXC NASA
contract NAS8-39073.  BE was supported by {\it Chandra} grants GO6-7008X 
and GO6-7098X. 

\end{acknowledgements}

\bibliographystyle{apj}

\clearpage

\end{document}